# DISTANCES TO NEARBY GALAXIES: COMBINING FRAGMENTARY DATA USING FOUR DIFFERENT METHODS

Dragan Huterer[1], Dimitar D. Sasselov[2,3], & Paul L. Schechter[1,2]

## ABSTRACT

The primary distance indicators are established in our Galaxy and the Local Group. There are at least four different methods which give good distances: methods using proper motions, RR Lyraes, Cepheid variables, and Type II supernovae. However the data on independent distances is very fragmentary, due partly to nature and partly to technological limits. As a result the data are rarely put together in a consistent way; instead, the discussion of distance scales is often focused on one or two methods or on individual objects. Hence the question: what is the current situation with our *overall* knowledge of distances to the nearby galaxies? We try to answer this question by combining the fragmentary data from all four methods for fifteen objects: the galactic center, the globular clusters M2, M3, M4, M13, M22, M92, and 47 Tuc, the galaxies IC1613, M31, M33, M81, M100, and M101, and the Large Magellanic Cloud. We pay special attention to covariances among the different distance estimates.

This most complete combination to date shows that all four methods are consistent within their random and systematic uncertainties.

## 1. Introduction

The demand for accuracy in primary distance indicators, like the Cepheids and RR Lyrae stars, has increased steadily in recent years. Nevertheless, certain problems have remained unresolved and are a cause for concern – the systematic offset between Cepheid and RR Lyrae distances, and the issue of Cepheid zero point dependence on metallicity, among them (Walker 1992; Gould 1994; van den Bergh 1995).

---

[1] Massachusetts Institute of Technology, Physics Dept., Cambridge, MA 02139; E-mail: dh@space.mit.edu, schech@achernar.mit.edu

[2] Harvard-Smithsonian Center for Astrophysics, Cambridge, MA 02138; E-mail: sasselov@cfa.harvard.edu, pschechter@cfa.harvard.edu

[3] Hubble Fellow

The present paper is an effort to bring together for comparison and scrutiny four basic methods, drawing widely upon the literature. In particular, we try to lift some of the degeneracy in the Cepheid *vs.* RR Lyrae problem by introducing completely independent methods – proper motions and supernovae. In this respect, we take advantage of the large amount of proper motions data now available as a result of the concerted effort of K. M. Cudworth, R. J. Rees, R. C. Peterson and associates on globular clusters (Cudworth 1986; Rees 1992; Peterson *et al.* 1995).

The data on independent distances has a highly fragmentary character. Some of the incompleteness is natural – old systems, like globular clusters, have no Cepheids or Type II supernovae; most of the incompleteness is technical – the nearby galaxies are too far for our ability to measure proper motions. We see our task as that of constructing the Eiffel Tower shown on the frontispiece of the *Decalages vers le rouge et expansion de l'univers* proceedings (Balkowski & Westerlund 1977); we use four methods to support our structure, but there are missing parts, joints, even levels. On top of it, of course, is $H_0$.

Therefore we look for consistency in the comparison from the whole sample, rather than paying more attention to individual well-studied objects. Our goal is to study in a comprehensive but consistent manner the distribution of the distances these four methods provide to 15 objects in the vicinity of our Galaxy, including the galactic center [GC]. We have no strong preference for any particular calibration or correction.

In the first section we describe briefly each of the four methods for distance determination. The next section has the details of their application to each of the 15 objects. This is followed by a discussion of the methodology and analysis of the results, ending with a summary.

## 2. The Methods for Measuring Distances

### 2.1. Combining Data from Different Methods

Before we describe each of the methods used, we outline our approach. We have $n$ galaxies and clusters for which we obtain distances. We shall represent those distances by an $n$-dimensional distance vector with an $n$-dimensional uncertainty ellipsoid. The coordinates of the center of the uncertainty ellipsoid will be best estimates of the distances to the objects. The uncertainty ellipsoid defines the error range of the distance vector.

Distance measurements are often correlated; we shall use covariance as an approximation of correlation between different measurements. The uncertainties of the distance measurements will be then represented by an uncertainty matrix $\mathbf{\Sigma}$, where the off-diagonal elements, $\sigma_{ij}{}^2$, are the covariance terms. The inverse of this matrix, the weight matrix $\mathbf{W} = \mathbf{\Sigma^{-1}}$, will give the coefficients of the equation of the uncertainty ellipse, or in our $n$-dimensional case – of the $n$-dimensional ellipsoid. The average distance vector, $\mathbf{v}_{ave}$, will then be:

$$\mathbf{v}_{ave} = (\sum_{i=1}^{N} \mathbf{W})^{-1} \sum_{i=1}^{N} \mathbf{Wv}, \tag{1}$$

where $\mathbf{v}$ is the $n$-dimensional distance vector and $N$ is the number of methods for estimating distances (we consider four, $N = 4$).

## 2.2. Dealing with Incomplete Data

The application of equation (1) is straightforward when there are measurements for all systems using all methods. But when a method has not been applied to one of the systems the corresponding element of the vector $\mathbf{v}$ is unknown. We take the expedient approach of choosing a value not unlike the other values for that system, but with a very large uncertainty. We adopt this as a bookeeping device, a placeholder, or what we prefer to call – a "wild guess". Thus, it is a "wild guess", yet perfectly correct, to estimate that your computer keyboard is $(50 \pm 45)cm$ wide. Similarly, if Cepheids give us a distance modulus to M31 of $24.5mag$, we can safely say that the proper motions distance modulus to M31 is $(20 \pm 10)mag$, although it has never been measured. In the statistical analysis the "wild guesses" will have vanishingly small weights with as large fractional error estimate as $10mag$ – they contribute very little to $\chi^2$. Also, we took the correlation between the "wild guesses" and measurements to any other galaxies to be zero as well for the sake of consistency and symmetry.

## 2.3. Choosing Data

In choosing data we used the following criteria: (1) absolute, not relative distances; (2) availability of more or less homogeneous distances; and (3) published results, not samizdat. We were willing to use preliminary results from conference proceedings, but we refrained from using (better quality) data which had been privately communicated but not yet published.

Our use of very incomplete methods perhaps needs explanation. Our reasoning is that any measurement using a new method adds additional information (and an additional check) which we are loath to forego.

### 2.4. Proper Motions and Kinematic Distances

Proper motions provide the simplest method, although they work only for the objects which are sufficiently close – ten out of fifteen objects that we examined have had internal proper motions measured. The basic idea is to measure the radial velocity **v** of the star, which, with suitable assumptions about isotropy, is a good approximation of the star's tangential velocity. The relationship between the tangential and angular velocity is $d = \frac{v}{w}$ where $d$ is the distance that we want to obtain.

Since the measurements of proper motions in different systems are uncorrelated, we have set all the covariances, $\sigma_{ij}$, equal to zero. This assumes that there is no systematic error associated with the assumed isotropy of radial and transverse velocities.

### 2.5. RR Lyraes

For the method of RR Lyrae variable stars, we have taken apparent magnitude $m$ (with its uncertainty $\Delta m$) as measured and correct it for reddening using the familiar term $\mathcal{R}E(B-V)$ and adopting for the ratio of total to selective absorption $\mathcal{R} = 3.1$, as suggested by many authors. (In some cases the correction for reddening was already made by the original author(s) and we didn't have to make it).

We calculated the absolute magnitude for RR Lyraes using the Carney *et al.* (1992a) relation

$$M = 0.15(\pm 0.01)[Fe/H] + 1.01(\pm 0.08)(mag), \tag{2}$$

which gives the absolute magnitude $M$ as a linear function of metallicity $[Fe/H]$. With a little calculation it is easy to see that the the biggest error in this equation is that of zero point, and that two other errors, ones from the slope term and metallicity term are much smaller and can be ignored. We therefore adopted $\Delta M = 0.08 \; mag$.

Now we can calculate the mean distance modulus as $DM = m_0 - M$. Since we have $\Delta m$ (random statistical error in $m$) and $\Delta M$ (systematic error in $M$), we can calculate the total uncertainty in distance modulus according to the usual formula for propagation of uncertainties $\Delta DM = \sqrt{(\Delta m)^2 + (\Delta M)^2}$. Calculations of distances using RR Lyraes have their errors scattered randomly, except for Carney's equation which is used with every object and which "pulls" the error in the same direction every time. Therefore, we have chosen to take $\sigma_{ij} = \sigma_{Carney} = 0.08$ as the covariance between any two objects.

We should note that Walker (1992) suggests a calibration, based largely on SN 1987A, and gives $M = 0.15[Fe/H] + 0.73 \ (mag)$ for which the zeropoint is almost 0.3 $mag$ brighter than Carney's zeropoint. However, more recent results seem to be closer to Carney's equation (Walker 1995).

### 2.6. Cepheids

For Cepheids, we have used the set of Caldwell & Coulson's (1987; henceforth CC) PL-V and PL-I equations (in the BVI$_C$ system). We did not use their PL-C relation, because it needs B-photometry in a significant way; such photometry is not available for any of the $HST$ Cepheid distances which we want to use. The equations for the PL-V and PL-I cases are

$$M_{PL-V} = -3.11\delta logP - 3.77 \quad \text{and} \quad m_{PL-V} = \langle V \rangle_0, \tag{3}$$

$$M_{PL-I} = -3.45\delta logP - 4.49 \quad \text{and} \quad m_{PL-I} = \langle I_C \rangle_0 \tag{4}$$

which are PL-V and PL-I (Kron-Cousins) relations for absolute and apparent magnitude respectively and where

$$\delta logP = logP - 0.9, \tag{5}$$

$$\langle B \rangle_0 = \langle B \rangle - \mathcal{R}_B E(B - V), \tag{6}$$

$$\langle V \rangle_0 = \langle V \rangle - \mathcal{R}_V E(B - V), \tag{7}$$

and

$$\langle I_C \rangle_0 = \langle I_C \rangle - \mathcal{R}_{I_C} E(B - V). \tag{8}$$

and $E(B - V)$ is reddening, $P$ is Cepheid's period of oscillation (in days); magnitudes with subscript "0" have been corrected for reddening. Here $\mathcal{R}_B$, $\mathcal{R}_V$ and $\mathcal{R}_{I_C}$ are ratios of total to selective absorption for the three bands respectively. We have adopted $\mathcal{R}_V = 3.1$ as used for most of our RR Lyrae distances. Following Taylor (1986) we obtain $\mathcal{R}_{I_C} = 1.77$, which includes the small term for G supergiants suggested by the author. We use the corrections for metallicity effects from CC in terms of *depletion*, defined as

$$d = 1 - 10^{[Fe/H]}. \tag{9}$$

For example, in the PL-V case this is the metallicity dependent bolometric correction

$$\Delta B.C. = (0.134BV_0 - 0.046)d \tag{10}$$

where

$$BV_0 = \langle B \rangle - \langle V \rangle - E(B - V). \tag{11}$$

Whenever B-photometry was not available, we used a relation with V-I from Caldwell & Coulson (1985).

The question of metallicity dependence trends in the Cepheid distances requires special attention (see Stothers 1988, and Gould 1994). We tried to test for metallicity dependence in the Cepheid distances in two different ways. First, by comparing them to the distances measured with RR Lyrae stars and minimizing the differences with a linear dependence on [Fe/H]. Second, by using the average distance ellipsoid of the entire sample to form these differences (see §4). In both cases we find a similar weak dependence on metallicity with a questionable goodness-of-fit to it.

We took data for $log\ P$, $\langle B \rangle$, $\langle V \rangle$, and $\langle I_C \rangle$ for a number of Cepheids in each object and calculated absolute and apparent magnitude for the PL-V, and PL-I case *for each Cepheid separately*. Whenever the observational data was in the Johnson system, we used the empirical relation obtained by CC from 45 Cepheids:

$$I_C = I_J + 0.272(V - I)_J - 0.049, \tag{12}$$

which is in good agreement with the general transformation in Taylor (1986). This concerns only the PL-I case. Then we obtained distance moduli $DM_{PL-V}$ and $DM_{PL-I}$ for each Cepheid. It was then easy to calculate average distance moduli for PL-V and PL-I case as the arithmetic means of all the corresponding distance moduli and their corresponding random uncertainties as

$$\Delta DM_{mean} = \sqrt{\frac{\sum_{i=1}^{K}(DM_i - \overline{DM})^2}{K(K - 1)}}, \tag{13}$$

where $\overline{DM}$ is the mean distance modulus and $K$ is the number of Cepheids in a galaxy that we considered.

Now we have PL-V and PL-I distance moduli with their corresponding random uncertainties. For a couple of our objects these moduli are almost identical, while the discrepancy seen in the rest is insignificant (§3). Therefore, we took the arithmetic mean as the final distance modulus and the arithmetic mean of the two uncertainties as the final random uncertainty, accounting for the above systematic difference in the uncertainties. We *did not* use the formula for propagations of uncertainties because PL-V and PL-I distance moduli are not independent at all. In our use of the CC PL-V and PL-I relations we have fixed the extinction laws (through the values of $\mathcal{R}_V$ and $\mathcal{R}_{I_C}$). However we have relied on Cepheid reddenings derived by Freedman *et al.* (1994a) using a method which is differential with respect to the LMC and may apply a different $\mathcal{R}_{I_C}$. We are not able to identify any origin for a systematic difference between the PL-V and PL-I distance moduli (see also Madore & Freedman (1991)), and only note that if any, the systematic is insignificant given the random uncertainties.

Feast (1991) notes that the CC distance scale is tied to van Leeuwen's parallax for the Pleiades, which is uncertain by $0.08\ mag$. Indeed, a preliminary $Hipparcos$ parallax for the Pleiades reported by M. Penston in the February 1994 $Observatory$ is $0.09\ mag$ more distant. Therefore, a good estimate for the covariances in the Cepheid method is $\sigma_{ij} = 0.08 mag$ for any two objects.

The final uncertainty is combination of random and systematic (covariance) uncertainty and we can calculate it using the usual formula for propagation of uncertainties $\Delta DM_{final} = \sqrt{\Delta DM_{mean}^2 + \sigma_{ij}^2}$.

In some systems Caldwell & Coulson gave a final uncertainty. However, while they used $\sigma_{ij} = 0.03$ as their systematic error, we prefer $\sigma_{ij} = 0.08$, leading to the following correction $\Delta DM_{final} = \sqrt{\Delta DM_{CC}^2 - 0.03^2 + 0.08^2}$.

## 2.7.   Type II Supernovae

The expanding photosphere method for SNe II was introduced by Kirshner & Kwan (1974) and is a variant of the Baade-Wesselink method used in radially pulsating stars (Baade 1926; Wesselink 1946).

The use of a distance correction factor derived from models of the expanding atmosphere (via fits to the synthetic spectral distributions) seems to divide the derived distances into shorter (Eastman & Kirshner 1989, Schmidt et al. 1994) and longer scales (Branch 1987, Baron, Hauschildt, & Branch 1994). The former scale is based on a physical, albeit simplistic, assumption of a two-level approximation, while the latter scale is an $ad$ $hoc$ corrected upwards parameter fit. We prefer to use the former scale.

## 2.8.   Remaining Inhomogeneities

Homogeneity and consistency are our main objectives in this study. We used each of the methods for deriving distances according to the prescriptions described above. Thus we avoided the use of distances derived by the various authors in order to retain control over the assumptions and uncertainties which enter into each distance determination. In a couple of cases we used data of inferior quality for the sake of preserving overall consistency. A case in point is our use of the optical V-band RR Lyrae relation for M4, instead of the superior new K-band relation which minimizes extinction problems. Despite our efforts, the fragmentary character of the data on independent distances leaves us with some remaining inhomogeneities.

In the kinematic distances the main problem with inhomogeneity arises because some of the proper motions and radial velocity data is not readily available. For example, we use a distance to M4 from Peterson et al. (1995) which is based on dynamical models by Rees (1995), while our distance to GC is simply $d = \frac{v}{w}$, where $v$ and $w$ come from separate studies. The kinematic distance to M13 is also model dependent (Lupton *et al.* 1987). Finally, the kinematic distance to the LMC is based on a very different approach altogether. All this is of no serious concern to us because all these kinematic distances share a common feature: they are completely independent from the other methods. Therefore our overall comparison of the four methods remains meaningful and consistent.

In the RR Lyrae distances there is one main inhomogeneity. In the globular clusters M2, M3, and M13 we used an extrapolation of the blue horizontal branch sequence, because of the lack of RR Lyrae stars or of their photometry. This reduces the precision but does not introduce a new systematic error.

In the Cepheid distances there is no particular inhomogeneity.

In the SNe II distances the inhomogeneity is caused by the fact that the SN 1987A in LMC and SN 1993J in M81 were not typical Type II supernovae and their distance correction factors may be systematically different (Schmidt *et al.* 1994).

## 3. Distance Calculations

### 3.1. The Galactic Center

Mould (1983) measured radial velocity for 49 M giant stars at the galactic center and obtained a dispersion $v = (113 \pm 11) km/s$. Spaenhauer et al. (1992) measured proper motions for giants with a dispersion $w = (0.29 \pm 0.02) \ arcsec/cent$. The distance is then $d = \frac{v}{w} = 8.21 \ kpc$. To calculate the uncertainty in $d$, we use the usual formula for propagation of uncertainties $\delta_d^2 = \delta_v^2 + \delta_w^2$, where $\delta_d = \frac{\Delta d}{d}$, $\delta v = \frac{\Delta v}{v}$ and $\delta w = \frac{\Delta w}{w}$. Therefore, $\Delta d = d\sqrt{\frac{\Delta v^2}{v} + \frac{\Delta w^2}{w}} = 0.98 \ kpc$. The corresponding distance modulus and uncertainty are then given by $DM = (14.57 \pm 0.26) mag$. It should be noted that if, as it now appears (e.g.

Blitz & Spergel (1991)), the bulge of the Milky Way is bar-like, our implicit assumption of isotropic orbits may not be correct.

Walker & Mack (1986) give, among other interesting results, complete data for 6 RR Lyrae stars in Baade's window near the Galactic Center (Table 1). Since we have $\langle V \rangle (= m_V)$, $[Fe/H]$ and $E(B - V)$ for each star, we can calculate their distance moduli separately, as described in the previous section. The mean value of those 6 distance moduli is $(14.29 \pm 0.08)$ $mag$. Taking into account the uncertainty (systematic error) in Carney's relation, we have $\sigma_{total} = \sqrt{\sigma_{mean}^2 + \sigma_{Carney}^2} = 0.11$ $mag$, giving $DM = (14.29 \pm 0.11)$ $mag$.

TABLE 1. Data for RR Lyraes in GC.

| star | $\langle V \rangle$ | $E(B - V)$ | $[Fe/H]$ | $\overline{DM}$ |
|------|------|------|------|------|
| 203 | 17.21 | 0.53 | -0.71 | 14.67 |
| 133 | 16.89 | 0.57 | -0.54 | 14.19 |
| 12 | 16.92 | 0.64 | -0.97 | 14.08 |
| 122 | 16.89 | 0.60 | -0.87 | 14.15 |
| 31 | 17.01 | 0.60 | -1.00 | 14.29 |
| 118 | 16.81 | 0.57 | -1.20 | 14.21 |

As regards a Cepheid distance, Caldwell & Coulson (1987) give us PL-C, PL-V and PL-I distance moduli for the GC which they obtained using the procedure explained in the previous section including all corrections. As noted above, we use only their PL-V and PL-I cases. Since they give $DM_{PL-V} = (14.42 \pm 0.18) mag$ and $DM_{PL-I} = (14.40 \pm 0.18) mag$, giving a mean distance modulus $DM = (14.41 \pm 0.18)$ $mag$. Correcting it for the covariance of $0.08$ $mag$ (as explained in the previous section), we find $DM = (14.41 \pm 0.19)$ $mag$. An implicit assumption in CC's distance is that orbits are roughly circular. If the potential of the Milky Way is as elliptical as Kuijken & Tremaine (1994) claim on the basis of the local velocity ellipsoid, CC have overestimated the distance.

No supernovae have been observed towards the GC, so we adopt a "wild guess" $DM = (14.00 \pm 10.00)$ $mag$.

### 3.1.1. Adding a new distance indicator

We decided to include the new and widely recognized method of measuring proper motions in the Galaxy (the water maser (WM) method) and see its impact on our average ellipsoid. Unfortunately, only one distance is available – the distance to the GC. Reid (1993) gives $d = (7.2 \pm 1.3) kpc$ which is equivalent to $DM_{GC} = (14.29 \pm 0.39) mag$. All the other distances in the WM distance vector will be our "wild guesses". Since Reid also specifies that the systematic error in the measurement to GC is $1.1 kpc$, we adopt, after conversion to magnitudes, that $\sigma_{xy} = 0.33 mag$ for any two elements in the WM uncertainty matrix.

After looking at any graph which contains the WM method or calculating the coordinates of a center of our new average distance ellipsoid, we can conclude that the impact of WM is absolutely negligible. The explanation is that the WM measurement has a very large uncertainty, and therefore, it has very little contribution in forming the average distance ellipsoid.

### 3.2. M 4

This is a low concentration globular cluster which is relatively near, but is heavily reddened by the Sco-Oph dark nebulosity with unusual extinction properties.

Proper motions and new radial velocities for nearly 200 stars in M 4 are combined in a statistical parallax distance measurement by Peterson, Rees, & Cudworth (1995). They obtain $d = (1.72 \pm 0.14) \; kpc$, so $DM = (11.18 \pm 0.18) \; mag$.

Most recently the RR Lyrae variables in M 4 have been studied by Liu & Janes (1990) – we used the light curves and individually determined E(B-V) for four of them. The metallicity of M 4 is $[Fe/H] = -1.05$ (Drake, Smith, & Suntzeff 1994), which confirmed independently the derivation of fundamental parameters for M 4 by Dixon & Longmore (1993), whose $\mathcal{R} = 4.0 \pm 0.2$ we adopt. The mean value of the four distance moduli and the systematic error in Carney's equation accounted for gives us finally $DM = (11.07 \pm 0.11) \; mag$.

There are no Cepheids and supernovae in M 4, so we use "wild guesses" $DM = (11.00 \pm 10.00) mag$.

### 3.3. M22

The globular cluster M 22 is similar to M 4 but farther away. It is similarly heavily reddened, but more uniformly and with a seemingly normal extinction law.

Peterson & Cudworth (1993) have measured proper motions of 672 stars in the field of M22, and thus obtained the mean distance $d = (2.57 \pm 0.29) kpc$, or $DM = (12.05 \pm 0.23) mag$.

Webbink (1985) gives $\overline{m} = 14.20$ $mag$, based mostly on photographic light curves of RR Lyraes by Wehlau & Hogg (1978), who preferred the value $\overline{m} = 14.1$ $mag$. We adopt the latter, as it agrees very well with the CCD photometry of M 22 by Anthony-Twarog *et al.* (1995). A reasonable guess for the uncertainty here is $\sigma_m = 0.09 mag$, which accounts also for the large uncertainty in the reddening $- E(B - V) = (0.42 \pm 0.03)$ by Crocker (1988). Lehnert et al.(1991) estimate $[Fe/H] = -1.54$. Now we have all the necessary data and, using the same procedure as before, we find $DM = (12.02 \pm 0.12) mag$.

There are no Cepheids or supernovae in M 22, so we just say that $DM = (11.00 \pm 10.00) mag$.

### 3.4.  47 Tuc

From Rees & Cudworth (1993) we have $d = (3.5 \pm 0.4) kpc$, or $DM = (12.72 \pm 0.24) mag$. This kinematic distance is uncertain because the globular cluster seems to have a significant differential rotation (Rees, 1995, private communication).

There is only one RR Lyrae star, V9, in this globular cluster. From Carney, Storm, & Williams (1993) we have $\langle V \rangle (= m_V) = (13.725 \pm 0.01) mag$ and $E(B - V) = (0.04 \pm 0.02)$. Zinn & West (1984) give $[Fe/H] = -0.71$. With this data we find $DM = (12.70 \pm 0.09) mag$.

There are no Cepheids and supernovae in 47 Tuc, so we just say that $DM = (12.00 \pm 10.00) mag$.

### 3.5.  M13

From Lupton *et al.* (1987) we have $d = (6.5 \pm 0.6) kpc$, or $DM = (14.06 \pm 0.20) mag$. From Webbink (1985) we have $\overline{m} = (14.95 \pm 0.05) mag$ and $E(B - V) = (0.03 \pm 0.02)$. Zinn & West (1984) give $[Fe/H] = -1.65$. With this data we find $DM = (14.09 \pm 0.09) mag$. There are no Cepheids or supernovae in M 13, so we just say that $DM = (12.00 \pm 10.00) mag$.

### 3.6.  M3

From Cudworth (1979) we have $d = (9.6 \pm 2.6) kpc$, or $DM = (14.91 \pm 0.52) mag$. From Webbink (1985) we have $\overline{m} = (15.68 \pm 0.05) mag$ and $E(B - V) = (0.00 \pm 0.02)$. Zinn & West (1984) give $[Fe/H] = -1.66$. With this data we find $DM = (14.91 \pm 0.09) mag$. There are no Cepheids and supernovae in M 3, so we just say that $DM = (13.00 \pm 10.00) mag$.

### 3.7. M2

From Cudworth & Rauscher (1987) we have $d = (11.0 \pm 1.7)kpc$, or $DM = (15.21 \pm 0.31)mag$. From Webbink (1985) we have $\overline{m} = (16.05 \pm 0.05)mag$ and $E(B - V) = (0.02 \pm 0.02)$. Zinn & West (1984) give $[Fe/H] = -1.62$. With this data we find $DM = (15.22 \pm 0.09)mag$. There are no Cepheids or supernovae in M 2, so we just say that $DM = (15.00 \pm 10.00)mag$.

### 3.8. M92

This is a very well studied metal-poor globular cluster with almost no reddening along the line of sight. Rees (1992) gives the distance obtained by measurement of proper motions $d = (8.3 \pm 1.6)kpc$ which is equivalent to $DM = (14.60 \pm 0.38)mag$.

Recent $BV$ light curves of 7 RR Lyrae stars by Carney *et al.* (1992b) give the mean apparent magnitude $\langle V \rangle (= m_V) = (15.15 \pm 0.01)mag$. Zinn (1985) gives $[Fe/H] = -2.24$ and $E(B - V) = (0.02 \pm 0.02)$. Using our usual procedure we get that $DM = (14.42 \pm 0.08)mag$.

There are no Cepheids and supernovae in M92, so we use "wild guesses" $DM = (13.00 \pm 10.00)mag$.

### 3.9. Large Magellanic Cloud

There are no internal stellar proper motions measurements in the LMC, but the distance to SN 1987A measured using its ring is derived in the spirit of astrometric measurements. We include it below and thus the LMC is the only object in our study which has its distance derived with all four methods.

Panagia *et al.* (1991) calculated the distance to SN 1987A by comparing the angular size of its circumstellar ring with its absolute size obtained by another method. From there, they derived the distance to LMC to be $DM = (18.50 \pm 0.13)mag$. Recently Gould (1995) reexamined the measurement of the caustics in the ionized-emission light curves of the ring and derived an upper limit to the distance to LMC of $DM = (18.37 \pm 0.04)mag$, assuming also that SN 1987A lies 500 pc in front of the LMC center. If adopted as a distance, rather than an upper limit, it halves the uncertainty in the average (from all four methods) LMC distance. However, the average distance changes only slightly (by 0.007 $mag$), and none of the other distances in our sample are affected; the total $\chi^2$ value in the comparison of all four methods shows only a slight improvement. Therefore, in order to retain our uniform approach, the Panagia *et al.* distance is used throughout. This is not to diminish the importance of the unprecedented precision of Gould's distance, but to illustrate the "egalitarian" nature of our approach.

Walker (1992) gives data for 7 clusters with about 180 RR Lyrae stars total. We proceed in completely the same way as in the case of GC: we calculate distance moduli for the 7 clusters and their mean with its own statistical uncertainty (Table 2). Then we incorporate the uncertainty from Carney's equation and finally obtain that $DM = (18.23 \pm 0.09)mag$.

TABLE 2. Data for RR Lyraes in LMC.

| cluster | $\langle V \rangle$ | $E(B-V)$ | $[Fe/H]$ | $DM$ |
|---------|------|----------|----------|------|
| NGC 1466 | 19.33 | 0.09 | -1.8 | 18.31 |
| NGC 1786 | 19.27 | 0.07 | -2.3 | 18.39 |
| NGC 1835 | 19.37 | 0.13 | -1.8 | 18.23 |
| NGC 1841 | 19.31 | 0.18 | -2.2 | 18.07 |
| NGC 2210 | 19.12 | 0.06 | -1.9 | 18.21 |
| NGC 2257 | 19.03 | 0.04 | -1.8 | 18.16 |
| Reticulum | 19.07 | 0.03 | -1.7 | 18.22 |

For the Cepheids, Caldwell & Coulson (1987) find, using the same calibration as for GC, that the best estimate of distance to LMC is $18.45mag$. They do not give the uncertainty here, but they give the uncertainty in their paper from 1986 (where they say that $DM = (18.65 \pm 0.07)$ is the best estimate, using a different calibration). After making correction for the systematic error of $0.08mag$ instead of $0.03mag$ we obtain $DM = (18.45 \pm 0.10)mag$.

The distance derived to SN 1987A in the LMC with the Type II Supernovae method is $49 \pm 6kpc$ (Eastman & Kirshner 1989) or $DM = (18.45 \pm 0.28)mag$.

### 3.10. IC 1613

IC 1613 is too far away for its stars to have its proper motions measured, so we adopt a "wild guess" $DM = (23.00 \pm 10.00)mag$ .

Saha *et al.* (1992) give us the RR Lyraes mean apparent magnitude using Gunn's green filter, $\langle g \rangle = (24.90 \pm 0.10)mag$. In the same paper we find the foreground absorption for the green band which is $A_g = 0.07$, and we get the corrected apparent magnitude $\langle g \rangle_0 = 24.83\ mag$. Since Carney's equation gives $M_V$, we need to convert $\langle g \rangle_0$ to the visual apparent magnitude $\langle V \rangle_0$. We do it with the help of the transformation by Kent (1985): $g = V + 0.41(B - V) - 0.19$, which requires knowledge of (B-V) when only one of Gunn's filters has been used. Fortunately, a good estimate of the intrinsic color of the RR Lyrae stars is possible on the basis of their periods and metallicity. We obtain a range (B-V)=$0.33 - 0.39$, corresponding to a small uncertainty in the transformed $\langle V \rangle$ of $0.006\ mag$. For the metallicity of RR Lyrae stars in IC 1613 we use $[Fe/H] = -1.6$, following the suggestion by Saha et al.(1992) and the metallicity estimate for the red giant branch by Freedman (1988). Finally, we have $\langle V \rangle_0 = (24.88 \pm 0.10)mag$, and using our procedure for RR Lyraes, we obtain that $DM = (24.11 \pm 0.13)mag$.

We take the complete $BVRI$ data for 10 Cepheids in IC 1613 from Freedman (1988). We adopt $[Fe/H] = -1.3$ (which gives a depletion $d = 0.95$), and $E(B - V) = 0.04$, both from Freedman's paper. Then we use the CC calibration as described above and get individual PL-V and PL-I distance moduli. The mean PL-V and PL-I distance moduli are $DM_{PL-V} = (24.35 \pm 0.16)mag$ and $DM_{PL-I} = (24.35 \pm 0.18)mag$ and our average distance modulus is $DM = (24.35 \pm 0.17)mag$. When we include the systematic error of $0.08mag$, we get that the final distance modulus is $DM = (24.35 \pm 0.18)mag$.

No supernovae have been observed in IC 1613, so we adopt a "wild guess" $DM = (23.00 \pm 10.00)\ mag$.

### 3.11. M33

Greenhill *et al.* (1993) were successful in measuring proper motions for five maser components in M 33 based on VLBI maps. The dispersions in transverse and radial velocities for the maser features provide a measure of consistency rather than a formal estimate of distance, which is $d = (600 \pm 300)kpc$, or $DM = (23.9 \pm 1.5)mag$.

For the RR Lyrae variables Pritchet (1988) gives $\langle B \rangle = 25.79$, $A_B = 0.31$, and $\langle B - V \rangle = 0.26$. Then we calculate the corrected visual apparent magnitude $m_V = \langle B \rangle - A_B - \langle B - V \rangle = (25.22 \pm 0.15)mag$. Mould & Kristian (1986) give $[Fe/H] = -2.2$, and then, using Carney's equation, we finally get $DM = (24.54 \pm 0.17)mag$.

We used complete data for 11 Cepheids in M33 from Freedman *et al.* (1991). In the same paper we can find that $E(B - V) = 0.10$, and a reasonable estimate for the metallicity $[Fe/H] = -0.3$, in view of the abundances for 42 supernova remnants in M 33 (Smith *et al.* 1993). Now, using our procedure for Cepheids, we get PL-V and PL-I distance moduli for each Cepheid. The mean PL-V and PL-I distance moduli are $DM_{PL-V} = (24.90 \pm 0.11) mag$ and $DM_{PL-I} = (24.87 \pm 0.10) mag$, and the average distance modulus is, therefore, $DM = (24.88 \pm 0.11) mag$. The final distance modulus is, after taking into account the systematic error, $DM = (24.88 \pm 0.14) mag$.

No supernovae have been observed in M 33, so we adopt a "wild guess" $DM = (23.00 \pm 10.00) \ mag$.

### 3.12. M31

Since M31 has no kinematic distance, we adopt $DM = (23.00 \pm 10.00) mag$. For a RR Lyrae distance Pritchet & van den Bergh (1987) give $\langle B \rangle = (25.68 \pm 0.06) mag$, $A_B = 0.31$, $\langle B-V \rangle = 0.26$ and $[Fe/H] = -0.6$. The visual apparent magnitude corrected for reddening is then $m_V = \langle B \rangle - \langle B-V \rangle - A_B = (25.11 \pm 0.06) mag$, and the distance modulus is, after using Carney's equation, $DM = m_V - M_V = (24.19 \pm 0.10) mag$.

We took Cepheid data for $\log P$, $\langle B \rangle$, $\langle V \rangle$ and $\langle I \rangle$ from the graphs given for three of Baade's Fields (fields I, III and IV) in Freedman & Madore's (1990) paper. In the same paper, we find that the metallicities $(\frac{Z}{Z_\odot})$ are 1.70, 1.14 and 0.30 and the reddenings $E(B-V)$ are 0.20, 0.25 and 0.00 for the three fields respectively. Then we calculated average PL-V and PL-I distance moduli for each field using the CC equations (Table 3). We took the average (arithmetic mean) of the 6 distance moduli that we thus obtained and the average of their uncertainties (we did not make any mistake by doing the latter thing, because the uncertainties were almost the same). The average distance modulus thus obtained is $DM = (24.38 \pm 0.13) \ mag$. After taking into account the systematic error of 0.08 $mag$, we get the final distance modulus $DM = (24.38 \pm 0.15) \ mag$.

TABLE 3. Mean distance moduli for the three Baade fields in M 31.

| window | $DM_{PL-V}$ | $DM_{PL-I}$ |
|--------|-------------|-------------|
| I | $24.28 \pm 0.14$ | $24.33 \pm 0.14$ |
| III | $24.36 \pm 0.15$ | $24.40 \pm 0.14$ |
| IV | $24.48 \pm 0.09$ | $24.40 \pm 0.11$ |

No supernovae have been observed in M 31, so we adopt a "wild guess" $DM = (23.00 \pm 10.00) \ mag$.

### 3.13.   M81

With no kinematic distance measured to M81 and no RR Lyrae stars known there, we adopt $DM = (27.00 \pm 10.00)mag$ as our "wild guesses".

We used data for the 25 Cepheids found and observed (Cousins V and I) by Freedman $et\ al.$ (1994a) with the $HST$. Metallicities for the M81 fields are only available from studies of HII regions (Garnett & Shields 1987; Zaritsky $et\ al.$ 1994) and give [Fe/H]≈0.05. The reddening is estimated by Freedman $et\ al.$ (1994a) to be E(B-V)=0.03. Then we calculated average PL-V and PL-I distance moduli from all the Cepheids using the CC equations, as applicable to VI data. The mean distance moduli are $DM_{PL-V} = (27.98 \pm 0.09)mag$ and $DM_{PL-I} = (27.95 \pm 0.08)mag$ and our average distance modulus is $DM = (27.97 \pm 0.12)\ mag$. When we include the systematic error of $0.08mag$, we get that the final distance modulus is $DM = (27.97 \pm 0.14)\ mag$.

The distance to SN 1993J in M81 was estimated by Schmidt $et\ al.$ (1993) at $2.6 \pm 0.4\ Mpc$ or $DM = (27.08 \pm 0.33)\ mag$.

### 3.14.   M101

Since M101 is too far away for proper motions observations and no RR Lyrae stars are known there, we adopt $DM = (29.00 \pm 10.00)mag$ as our "wild guesses".

We used data for the 4 Cepheids found and observed (BVRI photometry) by Alves & Cook (1995) at KPNO. Metallicities for the M101 fields are only available from studies of HII regions (Zaritsky, Elston, & Hill 1990), and are transformed by Alves & Cook assuming a solar [Fe/O] ratio to give [Fe/H] = $-0.28$. The reddening is estimated by the authors to be zero. Then we calculated average PL-V and PL-I distance moduli from all the Cepheids using the CC equations. The mean distance moduli are $DM_{PL-V} = (29.26 \pm 0.17)mag$ and $DM_{PL-I} = (29.45 \pm 0.19)mag$ and our average distance modulus is $DM = (29.36 \pm 0.25)\ mag$. When we include the systematic error of $0.08mag$, we get that the final distance modulus is $DM = (29.36 \pm 0.27)\ mag$.

The distance to SN 1970G in M101 is estimated at $7.4 \pm 1.5\ Mpc$ or $DM = (29.35 \pm 0.40)\ mag$ (Schmidt $et\ al.$ 1994).

### 3.15.   M100

Since M100 is too far away for proper motions observations and no RR Lyrae stars are known there, we adopt $DM = (30.00 \pm 10.00) mag$ as our "wild guesses".

We used data from the $HST$ observations of Freedman $et$ $al.$ (1994b) for 20 of their Cepheids in M100. These time-averaged intensity-weighted V and I magnitudes, kindly provided to us by Freedman (1995, private communication), are preliminary. As a metallicity estimate for the M100 Cepheids we adopt a value of $(\frac{Z}{Z_\odot}) = 1.25$, derived from the abundances of HII regions (Zaritsky $et$ $al.$ 1994), assuming a solar [Fe/O] ratio. The reddening is estimated by Freedman $et$ $al.$ (1994b) to be E(B-V)=0.05. We calculated average PL-V and PL-I distance moduli from all the Cepheids using the CC equations, as applicable to VI data. The mean distance moduli are $DM_{PL-V} = (31.23 \pm 0.14) mag$ and $DM_{PL-I} = (31.31 \pm 0.14) mag$ and our average distance modulus is $DM = (31.27 \pm 0.20)\ mag$. When we include the systematic error of $0.08 mag$, we get that the final distance modulus is $DM = (31.27 \pm 0.22)\ mag$.

The distance to SN 1979C in M100 is estimated at $15 \pm 4\ Mpc$ or $DM = (30.9 \pm 0.6)\ mag$ (Schmidt $et$ $al.$ 1994).

## 4. The Results

### 4.1. Graphical Representation

Each of the 4 methods we used gives a set of fifteen distances and uncertainties which can be represented as an ellipsoid in a 15-dimensional space. Then we made projections of each such ellipsoid (including the average distance ellipsoid) onto the different planes defined by the axes with distances to the objects. We show the projections for five pairs of objects in Figures 1 and 2.

We derive these projections in the following manner. For a given pair $(x,y)$ of objects (and a given method) we create a 2-dimensional distance vector and a $2 \times 2$ uncertainty matrix by taking the $i$-th and $j$-th element from the 15-dimensional vector $\mathbf{v}$, and by taking the elements $\sigma_{ii}, \sigma_{jj}$, and $\sigma_{ij}$ from the $15 \times 15$ uncertainty matrix $\boldsymbol{\Sigma}$ (see §2. and Eqn.1). We thus form the uncertainty matrix $\boldsymbol{\Sigma}_{xy} = \begin{bmatrix} \sigma_x^2 & \sigma_{xy}^2 \\ \sigma_{xy}^2 & \sigma_y^2 \end{bmatrix}$ where $\sigma_x$ is the uncertainty in the $x$ direction, $\sigma_y$ is the uncertainty in the $y$ direction, and $\sigma_{xy}^2$ is a covariance term. Then, from the general equation of the uncertainty ellipsoid

$$\mathbf{v^T}(\mathbf{Wv}) = 1, \tag{14}$$

the 2-dimensional weight matrix $\mathbf{W} = \boldsymbol{\Sigma}_{xy}^{-1} = \begin{bmatrix} p & q \\ q & r \end{bmatrix}$ will give us the coefficients of the equation of the uncertainty ellipse $px^2 + 2qxy + ry^2 = 1$ and it is plotted on Figures 1 and 2. When dealing with a higher-dimensional confidence region a projection, not intersection, is used to a lower-dimensional space (*e.g.* Press *et al.* 1992), and to the extent that the confidence region is approximated by an ellipse on the plane, our projections are a relevant illustration to the analysis in the next section. Observing five of them in Figures 1 and 2, we can offer the following general remarks.

First, we note that all the average ellipses (projections of the average ellipsoid) are smaller than the original ellipses and are centered between them, which we expected to be the case. Second, we note that the "wild guesses" are represented by ellipses which are so elongated in corresponding (uncertain) directions that they look almost like two parallel lines (Fig. 1b,c). If the distance in both directions is a "wild guess", then we do not see the ellipse at all (Fig. 1a), because it is extremely big (with axes $\sim 10\ mag$ long in each direction).

All the distances (components of distance vectors) contribute in forming the components of the average distance vector, but their contributions are not the same. The limiting factor in distances' contributions are their corresponding uncertainties; distances with big uncertainties are almost "ignored" in forming the average distance, while those with small uncertainties play a major role there. Furthermore, if we make a change in the distance to one of the galaxies, all the distances of the average vector will be affected, not only the component in the direction of that galaxy (although that component will be affected the most, of course). The reason for that is that all the distances are correlated (except the proper motions distances), so the change in one direction will imply the change in all the other directions.

## 4.2.   $\chi^2$ and some additional comments

We can compare the obtained distributions of distances in several ways. First, we combine all 15 objects with their distances and uncertainties obtained with 4 methods for measuring distances, "wild guesses" included, and derive a best estimate, $\mathbf{v}_{ave}$, as already defined by Eqn.(1). We compare the distributions of all three methods to their average by calculating

$$\chi^2_{total} = \chi^2_{KD} + \chi^2_{RR} + \chi^2_{Ceph} + \chi^2_{SN}. \tag{15}$$

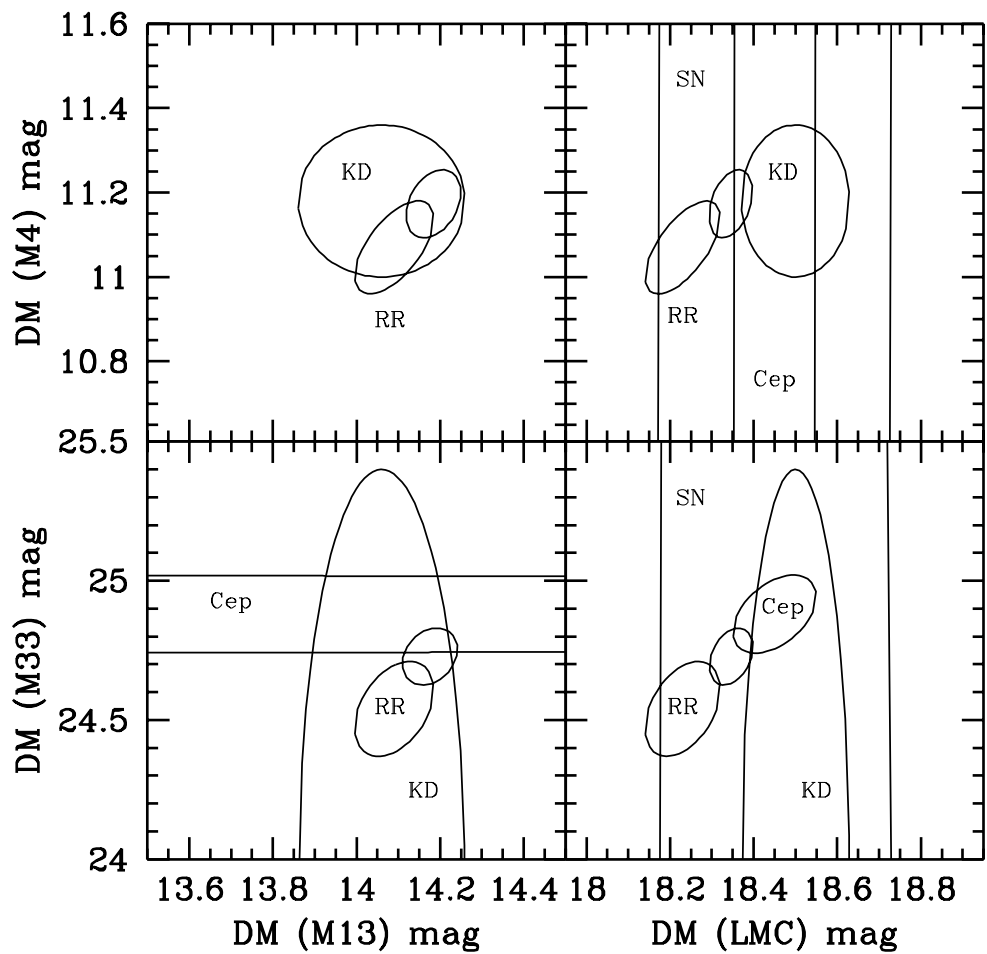

Fig. 1.— Projections of the distance ellipsoids for four objects of our sample. The ellipses due to each method are marked, *e.g.* KD is for kinematic distance; the small unmarked ellipses are the projections of the 15-dimensional average distance ellipsoid. There are no classical Cepheids in globular clusters, therefore "wild guesses" were used (see text), hence Cepheid distance ellipses are very elongated or not seen at all in such cases. The panels are marked (a) to (d), clockwise, from the upper left.

Here, *e.g.* for $\chi^2_{RR}$, the familiar $\chi^2$ equation:

$$\chi^2_{RR} = \sum_{k=1}^{15} \frac{(v_k - v_{ave})^2}{\sigma^2_k}, \tag{16}$$

becomes

$$\chi^2_{RR} = (\mathbf{v}_{rr} - \mathbf{v}_{ave})^{\mathbf{T}}[\mathbf{W}_{rr}(\mathbf{v}_{rr} - \mathbf{v}_{ave})], \tag{17}$$

where $\mathbf{v}_{rr}$ is the row matrix $\mathbf{v} = [v_1, \ldots, v_{15}]$. The $15 \times 15$ weight matrix of the measurements, $\mathbf{W}_{rr} = \mathbf{\Sigma}^{-1}_{rr}$, is not diagonal for the RR Lyrae method – the RR Lyrae distances are correlated and the covariance terms, $\sigma_{ij}$, (i,j=1,...,15), are not equal to zero (see § 2.).

In calculating $\chi^2_{total}$, the expected values, as represented by the average vector $\mathbf{v}_{ave}$, have not been adjusted or renormalized, hence the degrees of freedom are equal to the number of estimates, less the cases of "wild guesses" and parameters (15, in our case). We find $\chi^2_{total} = 13.0$ for 19 degrees of freedom; the four distance methods are generally consistent – Fig. 2 provides a limited illustration of this, as well.

An alternative method which avoids the use of "wild guesses", is to combine objects in two groups: kinematic distances and RR Lyrae, and RR Lyrae and Cepheids (see Table 4). In each group we derive a corresponding new $\mathbf{v}_{ave}$ and follow the procedure above (here the degrees of freedom will equal the number of objects in each group). In the comparison of RR Lyrae and Cepheids we have 5 objects and find $\chi^2 = 4.5$ for 5 degrees of freedom. For the combination of kinematic distances and RR Lyrae we have 9 objects and find $\chi^2 = 4.0$ for 9 degrees of freedom. A similar comparison between Cepheids and SNe II distances involves only 4 objects and gives $\chi^2 = 6.6$ for 4 degrees of freedom. Finally, we can compare Cepheids and kinematic distances, alas for 2 objects only, and find $\chi^2 = 0.31$ for 2 degrees of freedom – Fig. 2 is perhaps more informative in this case. A different way to present our approach here is to compare directly the distance distributions of the pairs, accounting for their combined uncertainties. The outcome is necessarily the same; we do it for illustration. Then for the comparison of RR Lyrae and Cepheids we will have:

$$\chi^2_{RR,CE} = (\mathbf{v}_{ce} - \mathbf{v}_{rr})^{\mathbf{T}}[(\mathbf{W}_{ce,rr})(\mathbf{v}_{ce} - \mathbf{v}_{rr})], \tag{17}$$

where $\mathbf{W}_{ce,rr} = (\mathbf{\Sigma}_{ce} + \mathbf{\Sigma}_{rr})^{-1}$. The number of degrees of freedom and $\chi^2$ values are the same as above.

Finally, we wanted to test the effect of inclusion of "wild guesses" in our analysis of the whole set: we repeated the above pair comparisons using all 15 objects in all cases. The results were consistent with the above ones to within a few percent in the $\chi^2$ values. This confirms the usefulness of adopting "wild guesses" in a problem with scarce and fragmentary data, in order to take advantage of the entire sample and account fully for statistical dependencies.

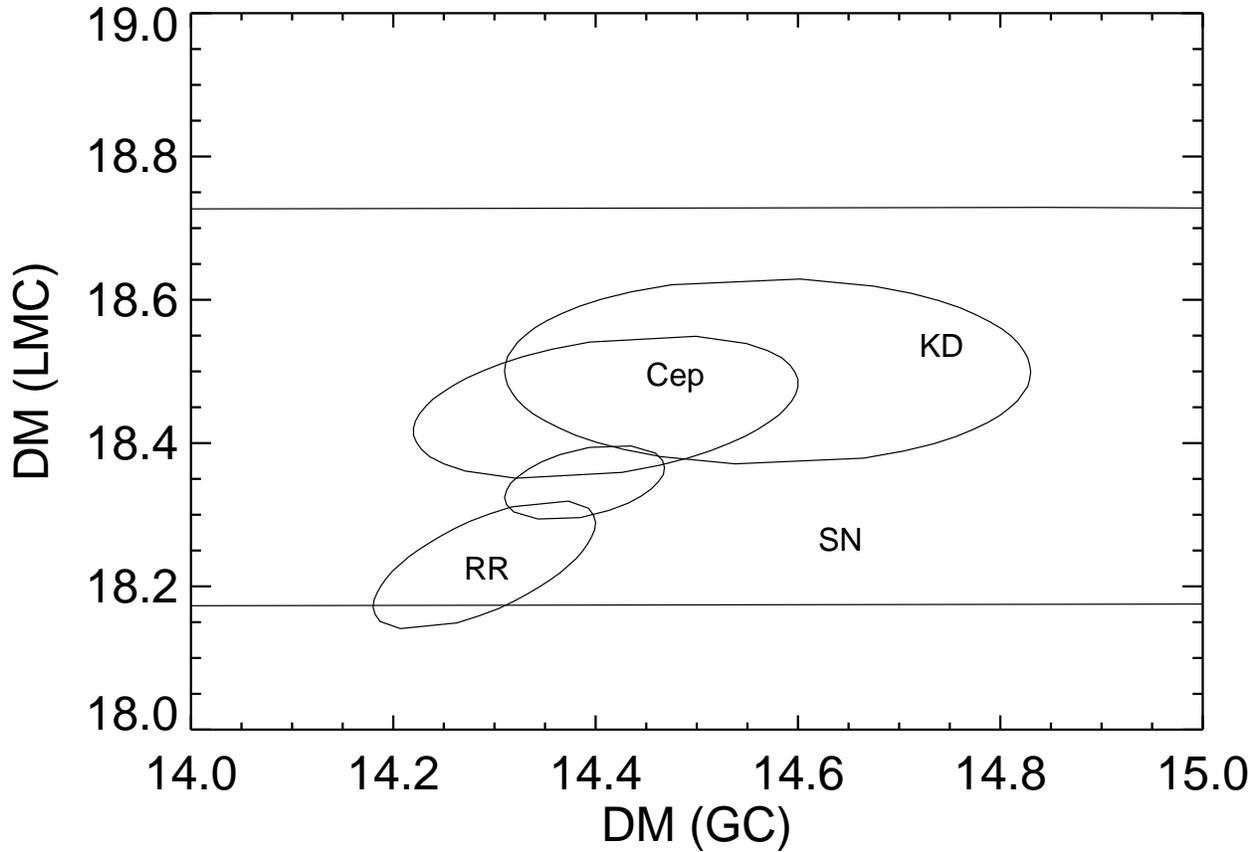

Fig. 2.— Projections of the distance ellipsoids for all four methods for the two objects for which these are available – the Galactic Center and the Large Magellanic Cloud. The unmarked ellipse is the projection of the average distance ellipsoid. Note that while for the LMC the kinematics and Cepheids agree better than with the RR Lyrae, for the GC it is the Cepheids and RR Lyrae which agree better.

We find a hint for a discrepancy in the RR Lyrae distances, with respect to correlated (Cepheids), as well as uncorrelated distance measurements. This is seen also on several of the graphical representations (Figs.1-2): RR Lyrae ellipses give smaller distance moduli ($\sim 0.1\ mag$) than the average ellipse. A possible reason for this slight disagreement of the RR Lyrae method with the other two could be the zeropoint in Carney *et al.* (1992a) equation for the absolute magnitude as a function of metallicity (see §2.5).

TABLE 4. Input Distance Moduli and Whole-Set Average.

| | $DM(KD)$ | $DM(RR)$ | $DM(Cep)$ | $DM(SN)$ | $DM(AVE)$ |
|---|---|---|---|---|---|
| $M\ 4$ | 11.18±0.18 | 11.07±0.11 | – | – | 11.17 |
| $M\ 22$ | 12.05±0.23 | 12.02±0.12 | – | – | 12.11 |
| $47Tuc$ | 12.72±0.24 | 12.70±0.09 | – | – | 12.79 |
| $M\ 13$ | 14.06±0.20 | 14.09±0.09 | – | – | 14.18 |
| $M\ 92$ | 14.60±0.38 | 14.42±0.08 | – | – | 14.51 |
| $GC$ | 14.57±0.26 | 14.29±0.11 | 14.41±0.19 | – | 14.39 |
| $M\ 3$ | 14.91±0.52 | 14.91±0.09 | – | – | 15.00 |
| $M\ 2$ | 15.21±0.31 | 15.22±0.09 | – | – | 15.31 |
| $LMC$ | 18.50±0.13 | 18.23±0.09 | 18.45±0.10 | 18.45±0.28 | 18.35 |
| $IC1613$ | – | 24.11±0.13 | 24.35±0.18 | – | 24.22 |
| $M\ 31$ | – | 24.19±0.10 | 24.38±0.15 | – | 24.28 |
| $M\ 33$ | 23.9±1.5 | 24.54±0.17 | 24.88±0.14 | – | 24.73 |
| $M\ 81$ | – | – | 27.97±0.14 | 27.08±0.33 | 27.79 |
| $M\ 101$ | – | – | 29.36±0.27 | 29.35±0.40 | 29.29 |
| $M\ 100$ | – | – | 31.27±0.22 | 30.9±0.6 | 31.15 |

## 5. Conclusions

To summarize, we analyze the most complete combination of Local Group distances to date. We compare four basic distance indicators by applying them to 15 objects – 7 globular clusters, 6 nearby galaxies, the galactic center, and the LMC. The globular clusters have no Cepheids and supernovae; five galaxies have no proper motions measured. All distances derived by the four methods to each of the 15 objects are compared and an average distance vector in this 15-dimensional space is derived. All values are listed in Table 4. We tested in several ways the statistical consistency of the distance distributions among each other.

Our conclusion from the $\chi^2$-tests and from the evidence in the projections of the distance ellipsoids is that all four methods provide distances which agree within their random and systematic uncertainties.

Finally, we offer a few comments. The point of our paper is in the consistent combination; the above conclusion comes from it. Pairwise comparisons of the methods show discrepancies, some of which had been well known. First, there remains a systematic discrepancy between the Cepheids and RR Lyrae, in that the latter distances are smaller. This needs to be combined with what appears to be a very good agreement between the kinematic and Cepheid distance measurements. While the former discrepancy is well known, the latter agreement is a new result, in the sense that there is only one object (the GC) where we could compare the Cepheid and kinematic distance scales truly directly; instead, here we could infer agreement from the whole sample. However, most kinematic distances being uncertain as they are, agree well with the RR Lyraes too, and the discrepant method cannot be identified unambiguously. In fact, with the SNe II scale being systematically smaller as well, it is the Cepheids which now appear discrepant (however, the SNe II method needs much further improvement to attain the same level of precision). Finally, we find no evidence for contradiction between the four distance scales, in the sense, suggested by van den Bergh (1995), that a future correction of a systematic in the RR Lyrae distances could cause discrepancy for some, while removing it for others. This means that the RR Lyrae scale could be zeroed to the Cepheids without destroying the overall consistency between the four methods found here. A side effect of this could be a decrease in the age estimates for globular clusters.

As an outlook for the near future, we feel that all four methods have plenty of room for improvement – not one of them stands out as being of much superior accuracy to the rest. We expect that the validity of our combined approach to extragalactic distances will endure because the expanding data set will always remain fragmentary.

*Acknowledgements*

We thank Yara I. Alma-Bonilla, without whose early efforts on this problem this paper might never have been written. Our special thanks to Bruce Carney, W. Freedman, A. Saha, Ruth Peterson and Richard Rees, for providing data in advance of publication and comments on the manuscript. We thank also J. Huchra, R. Kirshner, B. Schmidt, and R. Jones for their enlightening comments. We are grateful to John Caldwell for providing us with the data used in deriving equation 12. D.D.S. acknowledges support for this work by NASA through Hubble Fellowship grant HF-1050.01-93A awarded by the Space Telescope Science Institute, which is operated by the Association of Universities for Research in Astronomy, Inc., for NASA under contract NAS 5-26555.

## 6. Figure Captions

**Figure 1**:   Projections of the distance ellipsoids for four objects of our sample. The ellipses due to each method are marked, *e.g.* KD is for kinematic distance; the small unmarked ellipses are the projections of the 12-dimensional average distance ellipsoid. There are no classical Cepheids in globular clusters, therefore "wild guesses" were used (see text), hence Cepheid distance ellipses are very elongated or not seen at all in such cases. The panels are marked (a) to (d), clockwise, from the upper left one.

**Figure 2**:   Projections of the distance ellipsoids for all four methods for the two objects for which these are available – the Galactic Center and the Large Magellanic Cloud. The unmarked ellipse is the projection of the average distance ellipsoid. Note that while for the LMC the kinematics and Cepheids agree better than with the RR Lyrae, for the GC it is the Cepheids and RR Lyrae which agree better.